\documentclass[
 aip,
 amsmath,amssymb,
 reprint,
]{revtex4-1}

\usepackage{amsfonts}
\usepackage{graphicx}
\usepackage{dcolumn}
\usepackage{bm}

\usepackage[utf8]{inputenc}
\usepackage[T1]{fontenc}
\usepackage{mathptmx}
\usepackage{etoolbox}

\makeatletter
\def\@email#1#2{
 \endgroup
 \patchcmd{\titleblock@produce}
  {\frontmatter@RRAPformat}
  {\frontmatter@RRAPformat{\produce@RRAP{*#1\href{mailto:#2}{#2}}}\frontmatter@RRAPformat}
  {}{}
}
\makeatother

\usepackage{amsmath,amssymb,amsfonts,amsthm}
\usepackage{graphicx}
\usepackage{wrapfig}
\usepackage{xr}
\usepackage{hyperref}
    
\usepackage{cleveref}		
\usepackage{color}
\usepackage{xcolor}

\usepackage[ruled, linesnumbered]{algorithm2e}
\usepackage{booktabs}

\renewcommand*\arraystretch{1.3}

\usepackage[normalem]{ulem}
\newcommand{\stkout}[1]{\ifmmode\text{\sout{\ensuremath{#1}}}\else\sout{#1}\fi}

\graphicspath{{figures/}}

\newcommand\rev[1]{{\color{black}#1}}

\begin{document}
\preprint{AIP/123-QED}

\title{Reconstructing Kernel-based Machine Learning Force Fields with Super-linear Convergence}

\author{Stefan Blücher}

\affiliation{BIFOLD--Berlin Institue for the Foundations of Learning and Data, 10587 Berlin, Germany}
 \affiliation{Technische Universität Berlin, Machine Learning Group, 10587 Berlin, Germany}

\author{Klaus-Robert Müller}
 \email{klaus-robert.mueller@tu-berlin.de}
 \affiliation{BIFOLD--Berlin Institue for the Foundations of Learning and Data, 10587 Berlin, Germany}
 \affiliation{Technische Universität Berlin, Machine Learning Group, 10587 Berlin, Germany}

\affiliation{Korea University, Department of Artificial Intelligence, Seoul 136-713, Korea}
\affiliation{Max Planck Institute for Informatics, 66123 Saarbrücken, Germany}
\affiliation{Google Research, Brain team, Berlin, Germany}

\author{Stefan Chmiela}
\email{stefan@chmiela.com}
\affiliation{BIFOLD--Berlin Institue for the Foundations of Learning and Data, 10587 Berlin, Germany}
\affiliation{Technische Universität Berlin, Machine Learning Group, 10587 Berlin, Germany}

\date{\today}

\begin{abstract}
Kernel machines have sustained continuous progress in the field of quantum chemistry. In particular, they have proven to be successful in the low-data regime of force field reconstruction. This is because many equivariances and invariances due to physical symmetries can be incorporated into the kernel function to compensate for much larger datasets. So far, the scalability of kernel machines has however been hindered by its quadratic memory and cubical runtime complexity in the number of training points.
While it is known, that iterative Krylov subspace solvers can overcome these burdens, their convergence crucially relies on effective preconditioners, which are elusive in practice. 
Effective preconditioners need to partially pre-solve the learning problem in a computationally cheap and numerically robust manner. 
Here, we consider the broad class of Nystr\"om-type methods to construct preconditioners based on successively more sophisticated low-rank approximations of the original kernel matrix, each of which provides a different set of computational trade-offs. 
All considered methods aim to identify a representative subset of inducing (kernel) columns to approximate the dominant kernel spectrum. 
\end{abstract}
 
\maketitle

\section{Introduction}
In recent years, machine learning force fields (MLFF) have emerged as a valuable modelling tool in quantum chemistry \cite{noe2020machine, unke2021machine, keith2021combining, pinheiro2021choosing}. 
The promise of MLFFs is to combine the performance of classical FFs with the accuracy of computationally expensive high-level ab initio methods. 
In this setting, both, neural networks \cite{behler2007generalized,behler2011atom,frank2022so3krates,lubbers2018hierarchical,schutt2017quantum,unke2019physnet,unke2021spookynet,haghighatlari2022newtonnet,batzner20223_nequip} and kernel-based approaches \cite{rupp2012fast,Chmiela2018, chmiela2019sgdml, christensen2020fchl, deringer2021gaussian, faber2018alchemical, glielmo2017accurate, li2021efficient, sauceda2021bigdml} have been successfully applied. 
Kernel machines are generally considered to be more data efficient in modelling high-quality MLFFs~\cite{unke2021machine}, but yield large linear optimization problems when many training samples and/or large molecule sizes are involved~\cite{chmiela2022iterativesgdml} (e.g. in materials ~\cite{chen2017accurate, zhang2018strategy, sauceda2021bigdml} or biomolecules~\cite{unke2022accurate}). Linear systems are generally solved in closed-form, which has quadratic memory complexity. 
For large kernel systems, closed-form solutions are therefore not feasible and it is necessary to invoke iterative solvers, which do not require to store the full optimization problem at once~\cite{Gardner2018_GPyTorch, Liu_scalableGPs, Wang_oneMillionDataPoints}. Alas, iterative solvers are highly sensitive to the numerical properties of the kernel matrix: strong correlations within the data yield ill-conditioned optimization problems that are hard to converge, because small changes in the model parameters lead to vastly different responses\cite{chmiela2022iterativesgdml}. MLFFs are exposed to this issue due to predominantly stable atomic bonding patterns. This issue also occurs in models that incorporate differential constraints as inductive bias~\cite{schmitz2022algorithmic}.
Preconditioning techniques aim to diminish the strongest linear dependencies, by pre-solving parts of the system using a numerically more stable algorithm
and thereby separate the effects of strong correlations from the iterative solver. 
While many standard preconditioning approaches exist, effective solutions often rely on domain expertise.\\

In this study, we discuss the relevant theoretical and practical considerations necessary to develop the appropriate preconditioner for systems of many interacting atoms, in order to construct accurate MLFFs. An effective preconditioner has to represent all atomic cross-correlations faithfully, while being computationally cheap to construct.
Basic preconditioning approaches, such as Jacobi or sparse preconditioning, remove potentially important entries (correlations) from the kernel matrix and can therefore not reliably describe complex many-body interactions. The alternative is to use low-rank approximations that compress, but retain all entries in the kernel matrix. The optimal solution would be to find the most relevant subspace spanned by the kernel matrix via SVD decomposition, which is however prohibitively expensive in practice. 
A more economical approach is to approximate the kernel matrix using a subset of its columns as inducing columns. This is the key idea behind the Nyström method~\cite{Williams_nystromkernel} and the incomplete Cholesky decomposition~\cite{patel2016deterministic}, which differ in the way how inducing columns are sampled.  
These two methods are representative for a broad class of inducing point methods \cite{kumar2009sampling,musco2017recursive,patel2016deterministic} and offer opposing trade-offs with regard to construction cost and convergence speed.

\begin{figure*} 
    \centering
    \includegraphics[width=\linewidth]{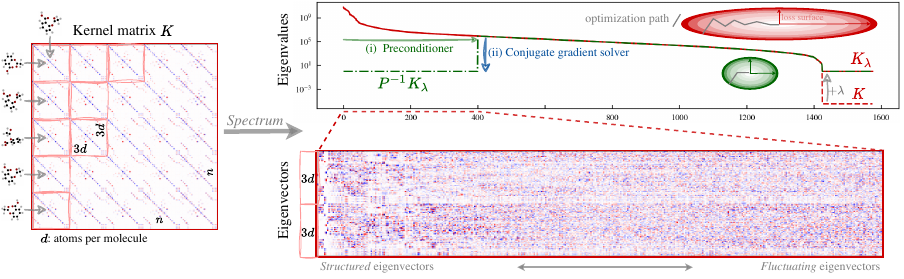}
    \caption{Left panel: sGMDL kernel for aspirin as a representative example for correlated molecular systems~\cite{chmiela2017machine}. 
    Right panel: A preconditioned iterative solver first removes the dominant spectral dimensions and then approximate the remaining spectrum increasingly accurately via a $\#$-dimensional Krylov subspace ($\#$ iteration steps).
    Large magnitude differences across eigenvalues lead to narrow valleys in the loss landscape (red ellipse), which obfuscates the optimization path. Preconditioning can remedy this by flattening the spectrum, leading to wider valleys (green roundish ellipse), which are easier to descent.
    To detail the dominant spectral structure, we focus on the first entries (specifying the contributions of the first two training molecules) of the dominant $\sim 400$ eigenvectors.
    }
    \label{fig:kernelmatrix}
\end{figure*}

We demonstrate in various numerical experiments, how these different preconditioning approaches affect the spectrum of the kernel matrix and hence the convergence of iterative solvers.
This allows us to derive a heuristic that helps practitioners to balance computational resources efficiently when reconstructing kernel-based MLFFs with iterative training algorithms.

\section{Scalable kernel solvers for quantum chemistry}  \label{sec:scalable_kernel_solvers}
\subsection{Kernel machines for MLFF} \label{subsec:mlff_kernels}
Kernel machines offer a powerful way to model complex functions and have successfully been applied to reconstruct MLFFs~\cite{rupp2012fast, Chmiela2018, chmiela2019sgdml, christensen2020fchl, deringer2021gaussian, faber2018alchemical, li2021efficient, glielmo2017accurate, sauceda2021bigdml}. 
Here, we focus on the symmetric gradient-domain machine learning (sGDML)~\cite{chmiela2019sgdml} model, which incorporates all important equivariances and invariances of molecular FF to be particularly data efficient~\cite{Chmiela2018}.
The atomic positions are encoded via a descriptor, to obtain a positive semi-definite (PSD) kernel matrix ${K}\in \mathbb{R}^{n\times n}$ with size $n = 3 \,d \,n_\mathrm{train}$ ($n_\mathrm{train}$ training molecules with $d$ atoms).
The resulting learning problem amounts to solving the regularized linear system
\begin{equation}
	K_\lambda \cdot \pmb{\alpha} = \pmb{y}
	\label{eq:linear_system}
\end{equation}
with the kernel $K_\lambda = K + \lambda I_n$ and atomic force labels $\pmb{y}$.
Using the solution $\pmb{\alpha}$, force predictions for new input conformations are queried at linear computational cost in the number of training points $n_\text{train}$. This makes kernel-based MLFFs cheaper than deep neural network architectures with comparable accuracy in many cases~\cite{chmiela2022iterativesgdml, houston2022permutationally}.
For example, consider the number of parameters for models trained with 1k aspirin examples in Ref.~\cite{frank2022so3krates}: 500k~(NewtonNet~\cite{haghighatlari2022newtonnet}) to 3M~(SpookyNet~\cite{unke2021spookynet}, NequIP~\cite{batzner20223_nequip}) in contrast to sGDML, which only used 63k parameters.

\Cref{eq:linear_system} is typically solved analytically in closed-form, which requires storage of the full matrix at a memory complexity of $\mathcal{O}(n^2)$ and cubic runtime cost $\mathcal{O}(n^3)$. Following this approach, memory complexity is the bottleneck in kernel-based methods, which prevents them to scale to large system sizes or numbers of training points.
Iterative solvers can overcome these limitations by solving the optimization problem numerically, using gradient descent. Gradient descent only relies on evaluations of matrix-vector products~\cite{chmiela2022iterativesgdml, freitas2005fast, hestenes1952methods, kim2005iterative, Gardner2018_GPyTorch, srinivasan2014preconditioned, rudi2017falkon, Wang_oneMillionDataPoints} and thus remedies the memory constraint, as no matrix needs to be stored explicitly~\cite{hackbusch1994iterative, saad2003iterative}.
Since $K$ is PSD, the linear system can be solved using the conjugate gradient (CG) algorithm, which is more efficient than plain gradient descent optimization~\cite{hestenes1952methods}.
The CG algorithm enforces conjugacy between optimization steps and thereby iteratively constructs a basis for the Krylov subspace of the kernel matrix with respect to the labels $\pmb{y}$.
This leads to a rapid progression towards the solution $\pmb{\alpha}$, with a theoretically guaranteed convergence in (at most) linear time $\mathcal{O}(\# \, n^2)$, with $\# \leq n$.
Depending on the spectral properties, i.e. the effective numerical rank of the kernel matrix, super-linear convergence in $\# \ll n$ is often achievable in practice~\cite{van1986rate,axelsson2000sublinear,beckermann2001superlinear}. In particular, the condition number (given by the ratio of largest to smallest singular value) and the eigenvalue decay-rate are the decisive properties that determine the overall performance of this algorithm.

\begin{figure}
    \centering

    \includegraphics[clip, trim= 1.5cm 0.6cm 0cm 2.1cm, width=0.15\linewidth]{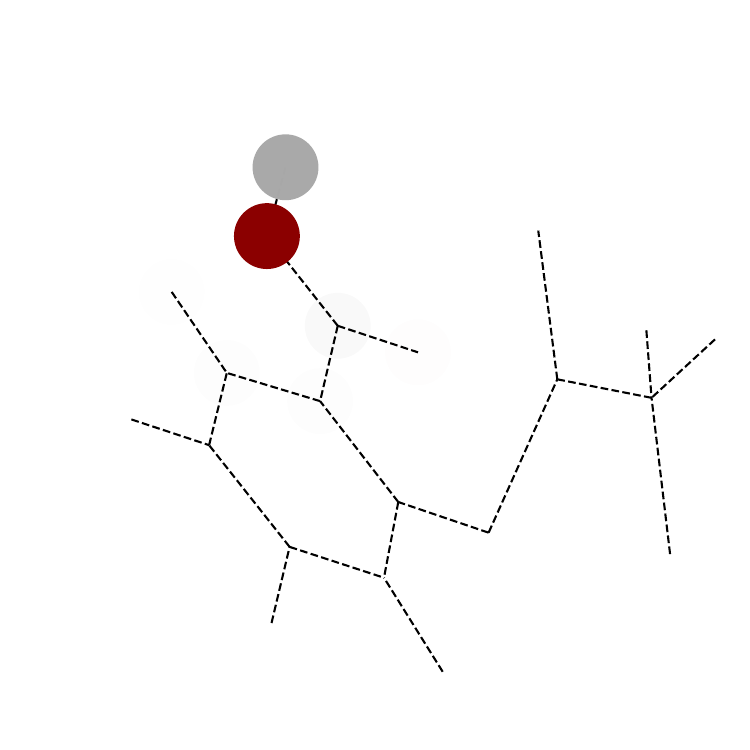}
    \includegraphics[clip, trim= 1.5cm 0.6cm 0cm 2.1cm, width=0.15\linewidth]{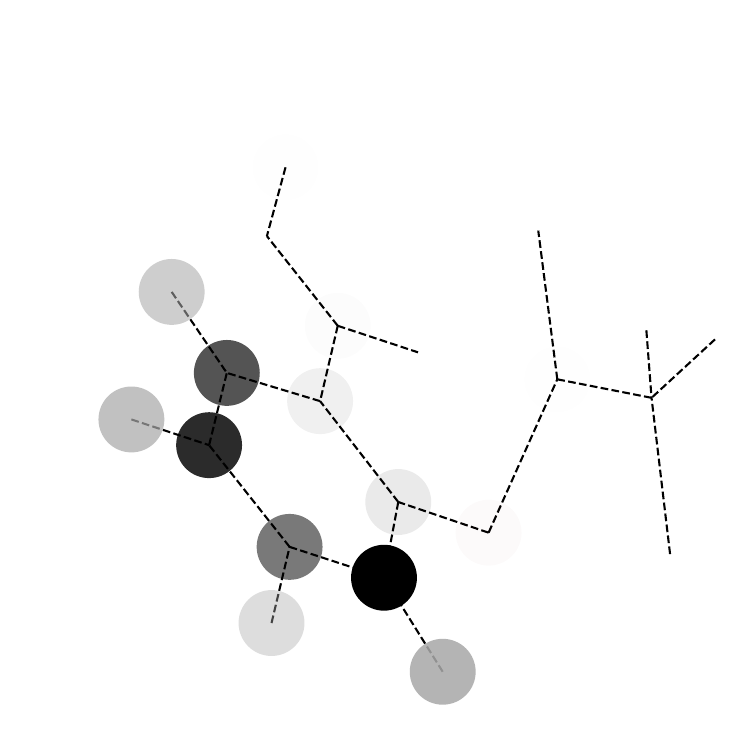}
    \includegraphics[clip, trim= 1.5cm 0.6cm 0cm 2.1cm, width=0.15\linewidth]{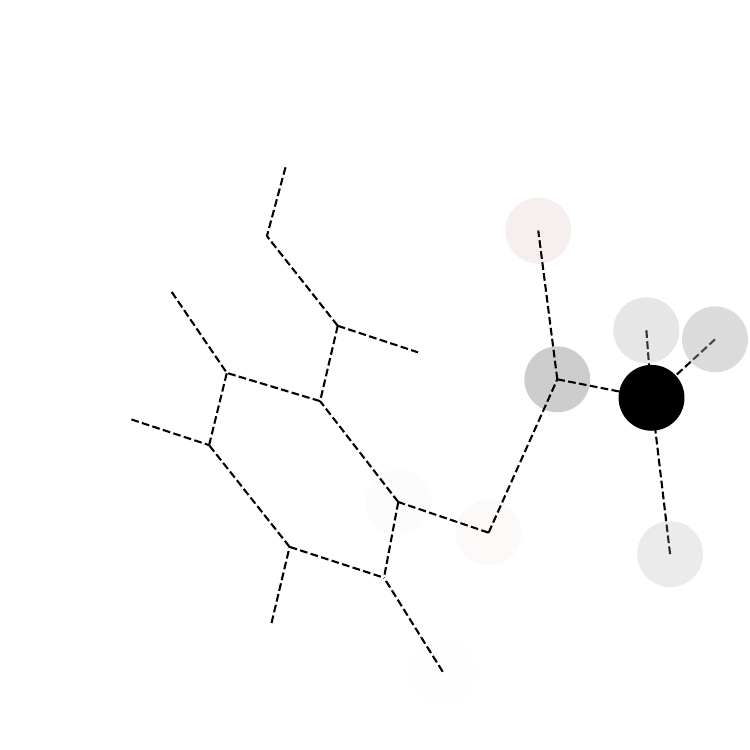}
    \includegraphics[clip, trim= 1.5cm 0.6cm 0cm 2.1cm, width=0.15\linewidth]{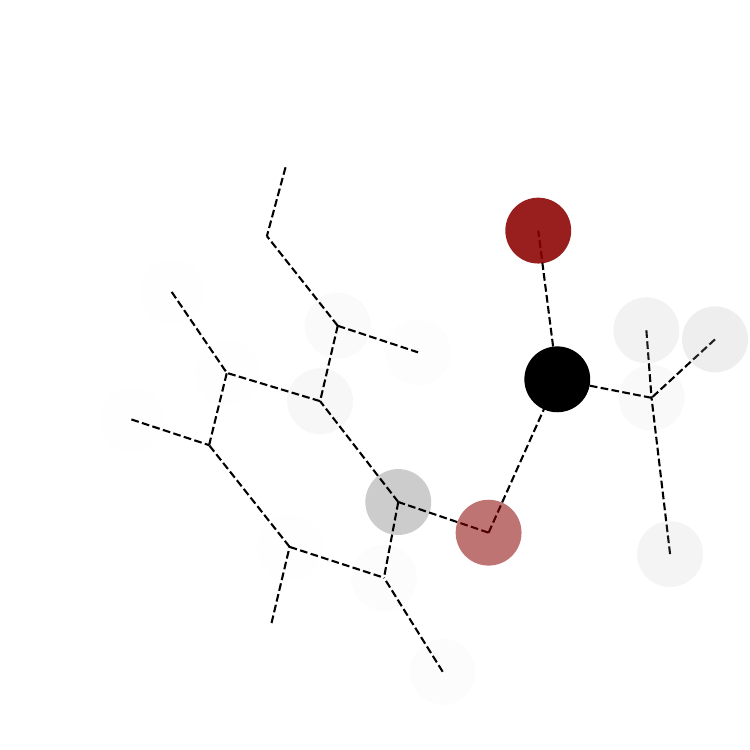}
    \includegraphics[clip, trim= 1.5cm 0.6cm 0cm 2.1cm, width=0.15\linewidth]{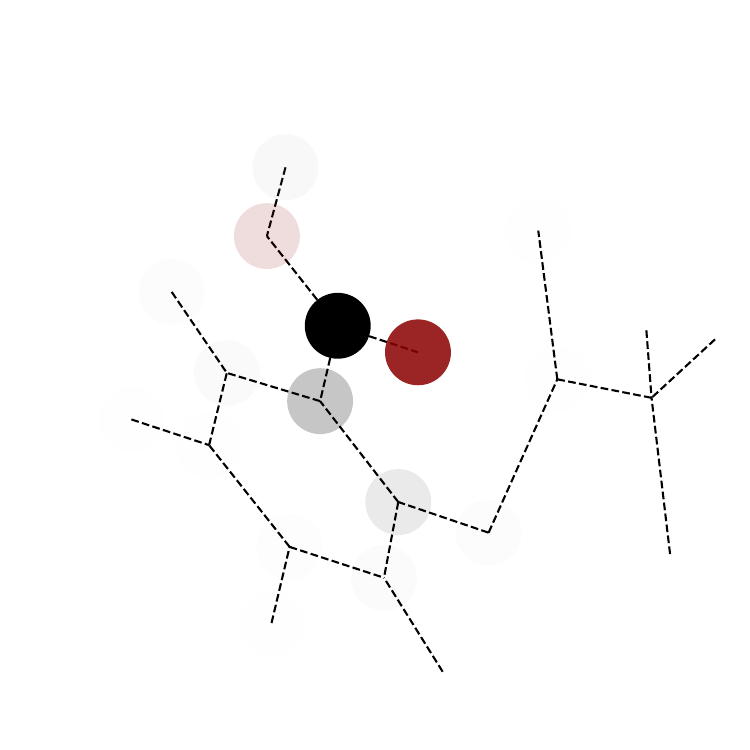}
    \includegraphics[clip, trim= 1.5cm 0.6cm 0cm 2.1cm, width=0.15\linewidth]{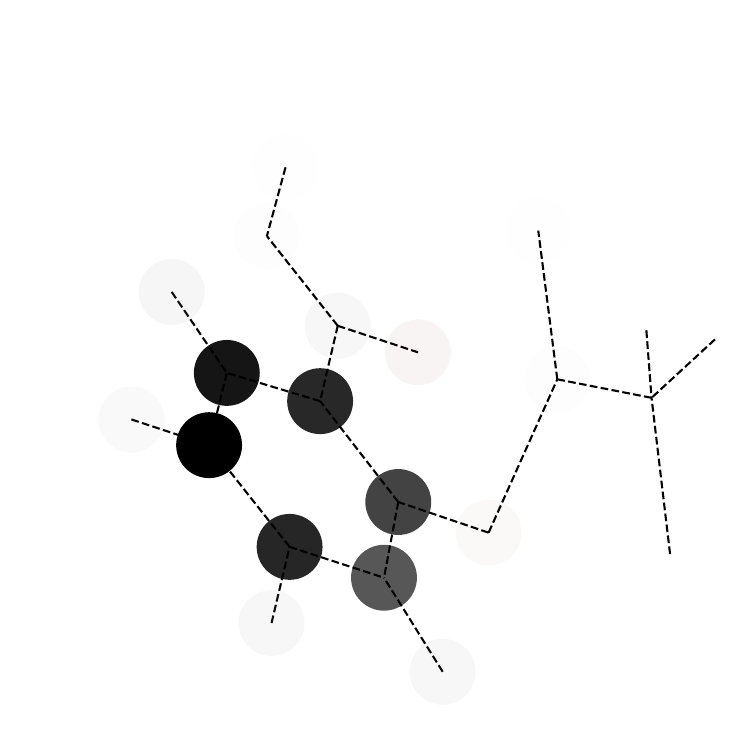}
    \caption{Atomic contributions per eigenvector (see \Cref{app-sec:technical_details}) via marginalizing the molecular periodicity and measuring the spatial L2 distance. 
    The kernel induces a low-rank decomposition, which forms the basis for a prototype molecule. }
    \label{fig:aspirin_eigenvectors}
\end{figure}
To illustrate this, we discuss these spectral properties for a specific example that is representative for the molecular datasets considered in this work (see \Cref{fig:kernelmatrix}, left panel) and most correlated systems in general.
The kernel matrix consists of $(3d \times 3d)$ block matrices, which correspond to correlations between pairs of training molecules. Each block represents the correlation structure between the $d$ atoms within each example. The molecular structures are restricted by the laws of quantum chemistry, which yield similar spectral characteristics across different systems. This enables us to derive transferable numerical insights about the kernel-based force field reconstruction problem. 
The eigenvectors and eigenvalues are shown in red (right panel of \Cref{fig:kernelmatrix}). For clarity, we only show the dominant 400 eigenvectors (columns) and their entries corresponding to two molecule (rows). The dominant eigenvectors exhibit a periodic pattern, which can be interpreted as a small set of prototypical molecular block structure elements. These patterns represent atom-wise contributions by each eigenvector (see \Cref{app-sec:technical_details} for details), which we visualize in \Cref{fig:aspirin_eigenvectors}. Intuitively, the structured eigenvectors form a low-rank basis for the full molecule, as they capture recurring sub-structures across all training points, e.g. strong correlations between interacting atoms.
On the other hand, the remaining low-magnitude eigendirections represent high-frequency fluctuations from the prototypical geometries. These eigenvectors contribute the individual geometric features of the training data~\cite{braun2008relevant}. This part of the spectrum decreases less rapidly, compared to the dominant part. Lastly, the spectrum also includes a (regularized) null space, which is unavoidable, due to a roto-translationally invariant representation of the molecule in terms of pairwise distances.

\subsection{Theoretical preconditioning for MLFF}
\begin{figure}
    \centering
    \includegraphics[width=\linewidth]{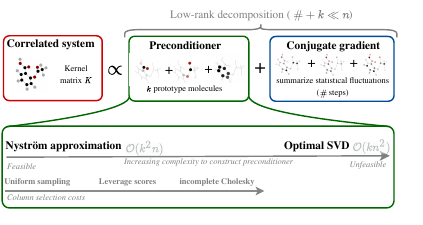}
    \caption{The preconditioned CG solver performs a numerically exact low-rank decomposition of the full kernel matrix $K$ (see \Cref{app-sec:technical_details} for details). Ideally, the preconditioner accounts for the correlations (prototypes) whereas the CG solver captures the remaining statistical fluctuations. When choosing the preconditioner we have to balance the trade-off between numerical costs and accuracy wrt. capturing more correlations. 
    }
    \label{fig:schematic_low_rank}
\end{figure}

Large differences in magnitude across eigenvalues (large condition number) induce long and narrow multi-dimensional valleys along the loss surface. These narrow valleys hamper gradient descent solvers, since the optimization path bounces back and forth on the way to the solution (as illustrated by the ellipses in \Cref{fig:kernelmatrix}). A suitable preconditioner is a full-rank matrix $P\approx K_\lambda$ that captures the dominant spectral components. It is used to normalize the matrix spectrum, which has the same effect as smoothening the underlying loss landscape \cite{avron2017faster, benzi2002preconditioning, cutajar2016preconditioning, kaasschieter1988preconditioned, ma2017diving, shabat2021fast, srinivasan2014preconditioned}.
With $P$, the linear system can equivalently be reformulated as
\begin{equation}
    P^{-1}K_\lambda \cdot \pmb{\alpha} = P^{-1}\pmb{y},
    \label{eq:preconditioned_linear_system}
\end{equation}
which preserves the original solution $\pmb{\alpha}$.
After applying the preconditioner, the remaining spectrum is then iteratively approximated within the Krylov subspace generated by the CG algorithm (blue arrow in \Cref{fig:kernelmatrix}). This two-step procedure can be regarded as a systematic low-rank decomposition of the kernel matrix (see \Cref{fig:schematic_low_rank} (top row) and \Cref{app-sec:technical_details}).

To retain the original PSD structure, we require that a symmetric decomposition $P = L L^T$ (${L}\in \mathbb{R}^{n\times n}$) exists \cite{avron2017faster,cutajar2016preconditioning, chmiela2022iterativesgdml}.
Another key requirement is that the construction of the preconditioner $P$ should not dominate the overall computational costs.
Generally, there are two principled ways to achieve this~\cite{zhou2011godec}: Either via sparsification of the kernel matrix (e.g. zero-out entries to obtain a block-diagonal form) or via a compression into a low-rank representation.
Only the latter approach is able to faithfully represent the complex correlation structure of quantum chemical systems.
In this work, we therefore focus on symmetric low-rank approximations $K \approx K_k = L_k L_k^T$ (${L_k}\in \mathbb{R}^{n\times k}$, $k \ll n$), which are expanded into invertible full-rank preconditioners after the regularization term is added:
\begin{equation} \label{eq:low_rank_LL}
     P = L_k L_k^T + \lambda I_n.
\end{equation}
The inverse $P^{-1}$ can then be computed economically via the Woodbury formula, 
\begin{equation} \label{eq:cholesky_precon}
    P^{-1} = \lambda^{-1}\left[I_n - L_{k} 
                        ( \lambda I_k + L_{k}^{\mathrm{T}}L_{k} ) ^{-1}
                                                                L_{k}^{\mathrm{T}} \right],
\end{equation}
under some numerical considerations~\cite{chmiela2022iterativesgdml}.
Overall, the runtime of this approach is $\mathcal{O}(k^2\, n)$ and it can be implemented at a memory complexity of only $\mathcal{O}(k\,n)$, by exploiting the symmetry of the matrix decomposition above \cite{Williams_nystromkernel}. 

The optimal, but also most expensive, rank-$k$ approximation $K_k$ is given by the singular-value decomposition (SVD), which is identical to the eigenvalue decomposition in this PSD matrix case.
A \emph{SVD preconditioner} can directly remove the $k$ dominant eigenvalues and hence optimally normalize the spectrum (green arrow in \Cref{fig:kernelmatrix}). 
Since, every eigenvector represents a unique linear combination of kernel columns, access to all columns is required to construct the SVD. Consequently, the computational cost of an SVD scales quadratically with kernel size, as $\mathcal{O}(k\,n^2)$. This scaling is prohibitive for large-scale MLFF reconstruction tasks and thus SVD is only considered as a theoretical bound for the optimal performance of rank-$k$ preconditioners in this work.

\section{Nyst\"om-type preconditioner} \label{sec:nystrom_precon_theory}

A low-rank approximation of the kernel matrix can also be constructed as a projection onto a \emph{subset} of its columns. This basic idea is underlying the Nyström method, which is traditionally used to approximate large scale kernel machines \cite{li2016fast, musco2017recursive, SUN_nystromreview, Williams_nystromkernel, smola2000sparse, drineas2005nystrom, zhang2008improved} and recently also as a preconditioner \cite{al2021two,cutajar2016preconditioning,frangella2021randomized,kim2018scaling}.
Given $k \ll n$ inducing columns $K_{nk} = \left(\begin{smallmatrix}
    K_{kk} \\ K_{(n-k)k}
\end{smallmatrix} \right)$, the Nyström approximation represents the full kernel matrix according to 
\begin{equation} \label{eq:nystrom_approximation}
\begin{split}
   K \approx K_k &= K_{nk} K^{-1}_{kk} K^\mathrm{T}_{nk} \\
        &= 
    \begin{pmatrix}
        K_{kk}      &   K_{(n-k) k}^T \\
        K_{(n-k) k}  &   K_{(n-k)k} K_{kk}^{-1} K_{(n-k) k}^T
    \end{pmatrix} \, .
\end{split}
\end{equation}
In this expression, the matrix dimensionality is implicitly encoded in the subscript, i.e., $K_{(n-k)k} \in \mathbb{R}^{(n-k)\times k}$. 
The $K_{kk}$ can be symmetrically factorized to obtain the form in \Cref{eq:low_rank_LL} (see Refs. \cite{chmiela2022iterativesgdml, foster2009stable} for implementation details).
The approximation error is given by the Schur complement $S = K / K_{kk} = K_{(n-k)(n-k)} - K_{(n-k)k} K_{kk}^{-1} K_{(n-k) k}^T$. Intuitively, all $(n-k)$ remaining columns are expressed in the basis of the inducing columns. Therefore, the accuracy is highly dependent on selecting the most representative set of inducing columns.

We now set out to find a set of expressive inducing columns that represent the dominant part of the spectrum and thus capture reoccurring structural patterns in the data, as depicted in \Cref{fig:aspirin_eigenvectors}.

\subsubsection*{Column selection strategies}
\textbf{Uniform sampling} The simplest approach is to draw a random sample of columns~\cite{Williams_nystromkernel, kumar2009sampling, Bach13_lowrank, cohen2015uniform}. This strategy is computationally inexpensive, but it essentially ignores the spectral properties of the kernel matrix and will therefore serve as a baseline in our study.

\textbf{Leverage score sampling} Not all columns contribute equally to the composition of the kernel matrix. To get a good approximation of the column space spanned by the kernel matrix, leverage scores can be used to measure how uniformly information is distributed among all columns~\cite{gittens2013revisiting,ipsen2014sensitivity, musco2017recursive, wang2013improving}. Columns that are strongly correlated, get assigned a low leverage score, while linearly independent columns mostly represent themselves and yield large leverage scores. Sampling according to leverage scores thus gives a higher chance of recovering a more representative set of inducing columns for the kernel matrix spectrum~\cite{braun2008relevant, alaoui2015fast}, although linear independence is not guaranteed. 

In our ridge regression setting, it is appropriate to consider \emph{ridge leverage scores} $\tau_i\big(K\big) = \left(K\left(K + \lambda I_n \right)^{-1}\right)_{ii}$~\cite{alaoui2015fast, mccurdy2018ridge}, which use a diagonal regularization parameter $\lambda$ to diminish small components of the kernel spectrum. Here, each kernel column is projected onto the regularized kernel to determine its overlap with the columns of the regularized kernel matrix. Since, evaluating this expression incurs the same computational cost as the original linear system, approximations are needed to make this computation feasible in practice~\cite{alaoui2015fast, chmiela2022iterativesgdml}.

\textbf{Incomplete Cholesky} Lastly, we discuss the \emph{incomplete Cholesky}~\cite{fine2001efficient, Gardner2018_GPyTorch, Wang_oneMillionDataPoints,foster2009stable} decomposition as a deterministic column selection strategy. The algorithm constructs a low-rank representation $K_k = L_k L^T_k$, by iteratively selecting inducing columns based on the Schur complement $S \in \mathbb{R}^{(n-1) \times (n-1)}$. 

In contrast to the two previous probabilistic approaches, the set of inducing columns is not selected at once, to avoid correlations. The Cholesky algorithm computes the residual Schur complement at each step and uses it as input for the next iteration, which then operates on the projected matrix with all previously selected inducing columns removed. In that sense, the Cholesky method uses successive Nyström approximations for each inducing column.
In the first iteration, we have (after permutation~$K_{n1} = \left(\begin{smallmatrix}
    K_{11} \\ \pmb{b}
\end{smallmatrix} \right)$):
\begin{equation}
    S = K / K_{11} = K_{(n-1)(n-1)} - \frac{1}{K_{11}} \pmb{b}\pmb{b}^T.
\end{equation}
The largest diagonal element of $S$ indicates the next inducing column, by which the approximation error is minimized according to the trace norm~\cite{Harbrecht2012}.
Since the trace is equal to the sum of all eigenvalues, this pivot rule focuses on the columns which are most representative for the remaining dominant spectral components in each Cholesky iteration. Despite being a greedy approach, the incomplete Cholesky systematically orthogonalizes correlated columns and thereby ensures linear independence, which enhances the representative power of the preconditioner~\cite{foster2009stable}. 
This leads close to optimal accuracy (comparable to SVD) in practise and theoretically guaranteed exponential convergence for exponentially decaying eigenvalue spectra\cite{Harbrecht2012}. 
Note that, if identical inducing columns are chosen for the incomplete Cholesky algorithm and the Nystr\"om method, both methods are equivalent \cite{kumar2009sampling, patel2016deterministic}.
The incomplete Cholesky algorithm is as costly as the overall Nyström preconditioner $\mathcal{O}(k^2\,n)$~\cite{Harbrecht2012} (i.e. it doubles the runtime costs). In contrast, the cost for the leverage scores estimation can be readily adapted by the fidelity of the approximation scheme~\cite{cohen2015uniform}.
Even though all three approaches are in the same complexity class $\mathcal{O}(k^2 n)$, their computational pre-factors are ordered according to (i) uniform sampling, (ii) leverage score sampling and (iii) incomplete Cholesky, also see \Cref{fig:schematic_low_rank} (bottom panel). 
We present an efficient incomplete Cholesky implementation in \Cref{app-sec:ICD}.

\section{Experimental results}

We use the MD17~(plus azobenzene)~\cite{chmiela2017machine} and MD22~\cite{chmiela2022iterativesgdml} benchmark datasets (available at \url{www.sgdml.org}) for our numerical experiments. Furthermore, we include a new catcher dataset, which consists of the buckyball batcher dataset in MD22, but without the fullerene (see \Cref{app-sec:experiments_all_molecules} for details).
We refer to these datasets in abbreviated form, by their molecule name, in this work (i.e. aspirin: MD17-aspirin dataset).

To establish a baseline, we first consider a small learning task ($n_\text{train}=250$ for aspirin), for which a comparision with the computationally expensive, optimal SVD preconditioner is still possible (Sections \ref{sec:experiment_spectrum} and~\ref{sec:experiment_cg_convergence}).
In the following \Cref{sec:experiment_rule_of_thumb}, we analyse the scaling behavior of all preconditioning approaches to large systems and derive a heuristic for preconditioner sizes that perform well in practice.  
Throughout our numerical experiments, we use the sGDML kernel~\cite{chmiela2019sgdml} with fixed length scale $\sigma=10$ and regularization $\lambda=10^{-10}$.

\subsection{Preconditioned spectrum} \label{sec:experiment_spectrum}
\begin{figure*}
	\centering
	
	\includegraphics[width=.99\linewidth]{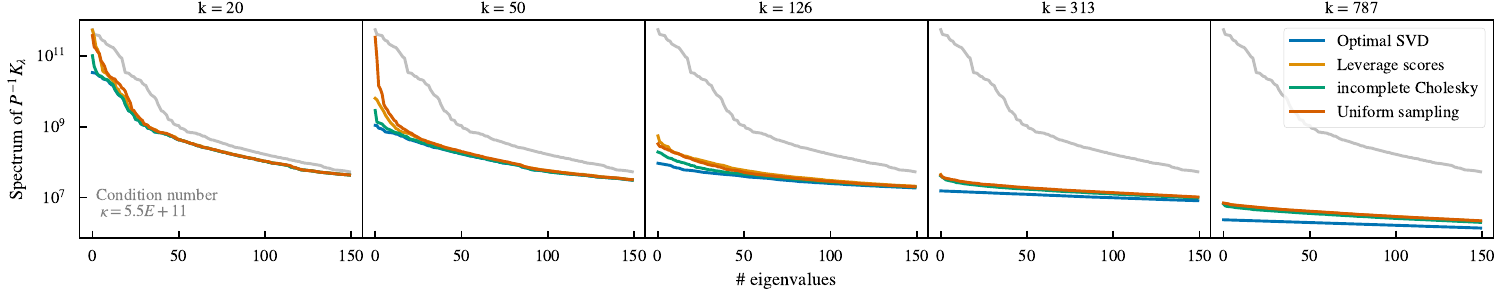}
	\caption{Preconditioned spectrum of $P^{-1}K_{\lambda}$ for varying preconditioner size $k$, restricted to the dominant 150 eigenvalues. Results for aspirin ($n_\mathrm{train}=250$, $n=15750$), more molecules in \Cref{app-fig:eigvals}. The grey curve corresponds to the original spectrum of $K_\lambda$. All Nyström-type approaches are lower-bounded by the optimal SVD preconditioner. 
	}
	\label{fig:eigvals}
\end{figure*}
We begin by analyzing the eigenvalue spectrum of the preconditioned kernel matrix $P^{-1} K_{\lambda}$, since the convergence rate of the iterative solver is directly connected to its condition number~\cite{van1986rate,axelsson2000sublinear,beckermann2001superlinear}. 
\Cref{fig:eigvals} shows the 150 dominant eigenvalues for aspirin for various numbers of inducing columns $k$. 
As expected from our theoretical considerations, SVD preconditioning is the most effective approach, leading to the smallest condition number, as compared to all Nyström-type preconditioning approaches considered in this work.
It is worth noting, that the optimal SVD preconditioner is never exceedingly better in this test (below one order of magnitude), independently of the number of inducing columns.

For small numbers of inducing columns, the incomplete Cholesky preconditioner shows a significantly better performance, compared to the two more basic probabilistic column sampling approaches.
The incomplete Cholesky decomposition removes the contributions of already sampled columns via the Schur complement and is therefore able to construct a more representative approximation of the spectrum using the same number of inducing columns.
In contrast, the inducing columns returned by the random or leverage score sampling approaches are potentially correlated, which can limit their effectiveness. As expected, the inducing columns generated by the incomplete Cholesky decomposition are not guaranteed to match the most important eigendirections, which makes it less effective than a SVD, yet cheaper to construct. 

Since the dominant eigenvectors are partially aligned with the kernel columns, the incomplete Cholesky preconditioner performs nearly optimally for a small number of inducing columns (up to $k \approx 3\,d$ in \Cref{fig:eigvals}). 
For larger preconditioners, this advantage diminishes. Here, the corresponding eigenvectors are not structured and do not significantly align with individual kernel columns anymore (see lower right panel in \Cref{fig:kernelmatrix}). 
Hence, we observe a constant performance gap with respect to the optimal SVD preconditioner with growing $k$.
In this scenario, the particular column selection is less relevant and the preconditioner size $k$ alone determines the approximation quality, despite a vastly varying construction cost between the different preconditioning approaches.
We observe a similar pattern for other molecular datasets in \Cref{app-fig:eigvals}.
For the sake of completeness, we also demonstrate the limitations of local Jacobi preconditioning in the same experimental setup (see~\cite{anzt2019adaptive,concus1985block,knyazev2001toward,lee2002performance,si2014memory} in \Cref{app-fig:spectrum_local_preconditioning}. \\

\subsection{Convergence speed of conjugate gradient} \label{sec:experiment_cg_convergence}
\begin{figure*}
    \centering

     \includegraphics[width=\linewidth]{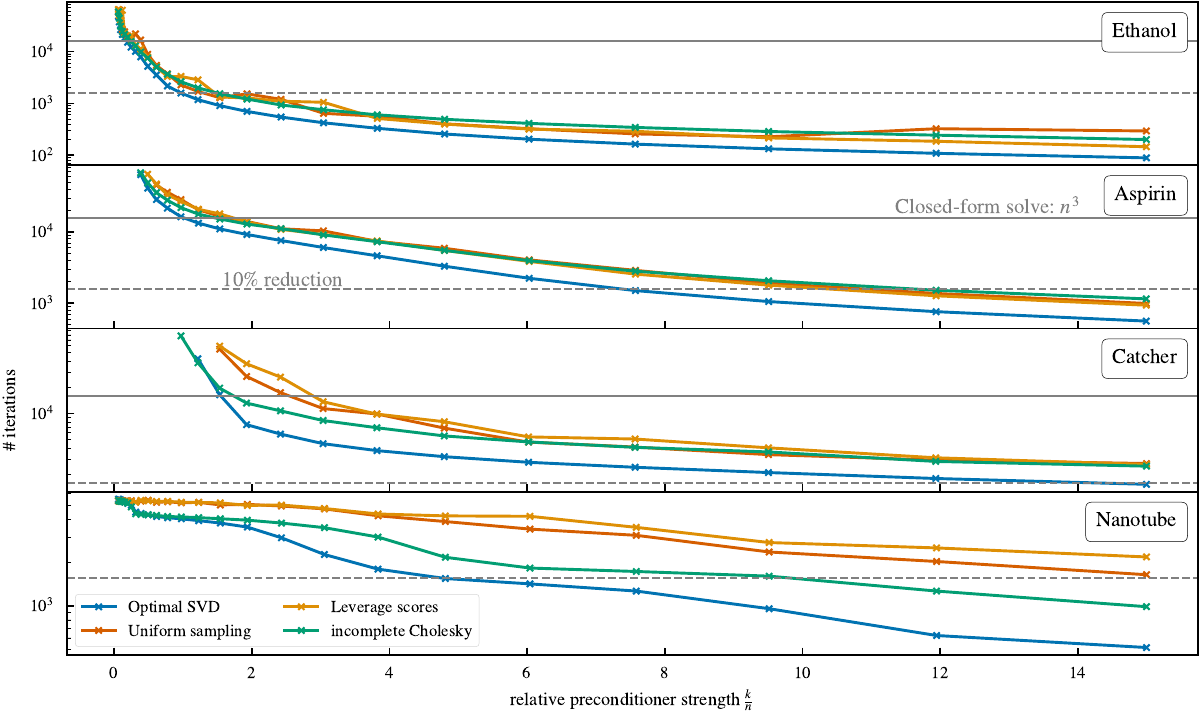}
	\caption{
 Iterative convergence speed for varying preconditioner strength for fixed kernel size $n\approx 15\text{k}$ ($n_\text{train} = 250$ for aspirin). 
 Only minor performance differences are visible for ethanol. In contrast, for more complex molecules (catcher, nanotube) the elaborate incomplete Cholesky preconditioner is superior to probabilistic sampling of inducing columns.
 }
	\label{fig:cgsteps}
\end{figure*}

Next, we quantify the iterative solver runtime by tracking the number of CG steps $\#$ until convergence, as this performance metric is insensitive to implementation and hardware details. 
\Cref{fig:cgsteps} shows the relationship between convergence speed and relative preconditioner strength $\frac{k}{n}$.
We consider molecules of different sizes, including ethanol ($d=9$), aspirin ($d=21$), a catcher molecule ($d=88$, see \Cref{app-sec:experiments_all_molecules}) and a large nanotube ($d=370$). Additional systems are shown in \Cref{app-fig:cgsteps_all_molecules}. 
The preconditioner strength is varied between $k\sim 50$ up to 14\% of all columns. 

The grey solid horizontal line corresponds to the cubic costs of the closed-form solution, the dashed line gives a visual cue for a $10\% \cdot n^3$ reduction, to provide a scale for relative computational complexity. For example, ethanol is simpler (easier to solve), compared to larger or more complex molecules, such as aspirin. 
We observe, that the functional dependency between preconditioner strength and computational cost differs between all molecules. 
In particular, the number of inducing columns necessary to achieve super-linear convergence depends on the complexity of the molecule. These different complexities are embodied within the characteristic shape of dominant part of the spectrum (see \Cref{app-fig:eigvals} for explicit spectra). 
For example, the catcher molecule requires the most inducing columns, since its spectrum contains many nearly degenerate, dominant eigenvalues. Similarly, it is difficult to accelerate convergence for the nanotube dataset, since the dominant part of the spectrum is only weakly decaying. 
In contrast, ethanol is simple to solve and more than 2\% preconditioning already leads to rapid convergence (one order of magnitude faster). 
Generally, undersized preconditioners lead to more CG steps, which incur a computational cost that is similar to that of analytic closed-form solutions. In constrast, more inducing columns generally lead to faster convergence, but at increased preconditioner construction cost (see \Cref{sec:experiment_rule_of_thumb} for practical guidance on this trade-off). 

Having discussed commonalities between all approaches, we now analyse their differences with respect to their applicability and numerical performance when applied to different molecules.
From \Cref{fig:cgsteps} it is clear that the SVD preconditioner is the most effective approach. As expected and already observe in \Cref{sec:experiment_spectrum}, it lower bounds all other Nyström-type preconditioners. 
We emphasize again, that the SVD preconditioner is too expensive in practise and only used here as the theoretical optimum. 
The performance gap between the optimal SVD preconditioner and all three Nyström-type preconditioners is largely determined by how well the inducing columns align with the eigenvectors. 
All three Nyström-type preconditioners perform similarly for most molecules, e.g, consider ethanol in \Cref{fig:cgsteps}, for which barely any difference is visible. This behavior is similar for to all additional molecules presented in \Cref{app-fig:cgsteps_all_molecules}.
Note, that the structural changes within the eigenvectors when applying larger preconditioners also \rev{degrades} the differences observed for the weakly preconditioned aspirin molecule (e.g. for $k<100$ in \Cref{fig:eigvals}).

This indicates that a uniform subset of columns is sufficiently expressive for the kernel spectrum and therefore no significantly better set of inducing columns can be found to further improve preconditioner performance. 
However, both probabilistic column selection approaches introduce fluctuation into the preconditioner performance, because any potential correlations between inducing columns are disregarded (i.e. they can draw a \emph{bad} set of inducing columns by chance. In contrast, the incomplete Cholesky performance is monotonously improving and does not fluctuate. This is reminiscent of the deterministic pivoting rule, which systematically improves the approximation error via repeated rank-1 updates at each iteration.

Further, we observe noteworthy differences for the larger and more structured catcher and nanotube molecules (two bottom panels in \Cref{fig:cgsteps}). 
Here, the incomplete Cholesky preconditioner significantly outperforms both probabilistic column sampling approaches. For small preconditioner sizes the incomplete Choleksy's performance is close the optimal SVD baseline. 
For the catcher molecule, this advantage diminishes and for larger preconditioners all Nyström-type approaches perform equivalently. 
For the nanotube molecule there are persistent differences between all preconditioner approaches.
Here, the performance gap between the SVD preconditioner and both probabilistic approaches increases with the number of inducing columns.
This indicates that both approaches do not construct an expressive set of columns. 
We attribute this to the symmetric, repetitive structure of the large nanotube, which creates many correlated (i.e. redundant) kernel columns. As outlined in \Cref{sec:nystrom_precon_theory}, such a scenario is problematic for probabilistic sampling, as it inevitably incorporates correlated inducing columns.
Here, the systematic selection of independent inducing columns by the incomplete Cholesky approach appears to play a crucial role to solely incorporate the relevant molecular structure.

Lastly, we note, that the leverage score sampling approach does not appear to significantly outperform the uniform sampling baseline, even for small preconditioners ($\sim 2\%$ or $k\sim 300$) (see \Cref{fig:cgsteps}). 
This is somewhat counter-intuitive as leverage scores are designed to improve on basic uniform sampling. 
We speculate that, this is related to the fact that the regularization $\lambda$ for the ridge leverage scores is fixed and not adapted to changing number of inducing columns $k$. Hence, ridge leverage scores are indicative for the spectral properties of the kernel matrix ($K_{\lambda}$), instead of directly targeting the dominant $k$-dimensional eigenspace ($K_k$), which is relevant for preconditioning.
We experimentally investigate this intuition in \Cref{app-fig:cgsteps_all_molecules}, via an improved (however computationally costly) notion of leverage scores. These improved leverages scores indeed remove the gap with respect to the uniform sampling baseline. 
However, they do not improve the preconditioner performance significantly. Therefore, we conclude that the incomplete Cholesky preconditioner is the most capable approach, as it is highly effective in selecting a representative set of inducing columns. This leads to nearly optimal preconditioning for a large, structured nanotube molecule. 
However, this relative advantage generally degrades with increasing number of inducing columns. For larger preconditioners it is sufficient to uniformly sample columns, since small eigendirections are less structured as they obtain support from most atoms simultaneously. 
To arrive at the right preconditioner choice, we additionally need to account for the different computational costs associated with each preconditioner (see also \Cref{sec:nystrom_precon_theory}). Since, the probabilistic preconditioners are generally cheaper, in many situations (e.g. large preconditioners or unstructured molecules such ethanol/uracil) random sampling should be preferred over the more capable but also more costly incomplete Cholesky approach.

\begin{figure*}
    \centering
	\def\arraystretch{0.9} 		
	\setlength\tabcolsep{8pt}		
    \begin{tabular}{cc}
        \hspace{1.4cm}Vary kernel size $n$ (Aspirin)    &       \hspace{1.3cm}Vary molecules ($n=31400$) \\
        \includegraphics[width=0.47\linewidth]{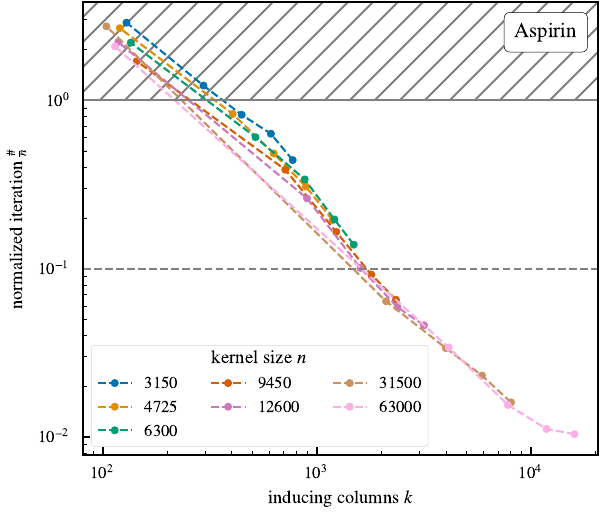} & 
        \includegraphics[width=0.47\linewidth]{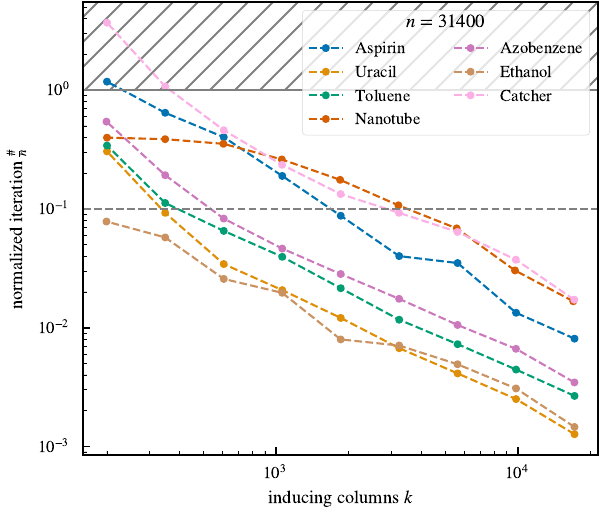}
    \end{tabular}
    \caption{
    The super-linear convergence ($\frac{\#}{n} < 1$) of iterative solvers is related to the preconditioning strength $k$ via a monomial power-law (linear dependence in the log-log space).  
    For increasing kernel size $n$ relatively less columns $k$ are needed to obtain an equivalent super-linear advantage. 
    Hence, preconditioning is increasingly valuable (left panel). 
    The monomial trade-off proportionality is verified for any molecule but the exact relation depends on the molecule type (right panel).}
    \label{fig:preconditioner_size}
\end{figure*}

\subsection{Optimal number of inducing columns} \label{sec:experiment_rule_of_thumb}
Next, we investigate the iterative convergence behavior, when scaling from smaller to larger kernel systems. 
To this end, we consider the normalized iteration $\frac{\#}{n}$, which measures the relative advantage of the preconditioned iterative solver compared to the closed-form solution. A normalized iteration of one indicates equivalent runtimes between both methods. 
For simplicity we restrict this analysis to leverage score-based preconditioning in the following, which lower bounds the performance of all other Nyström-type preconditioners. Here, the lower bound refers to the worst performance (a better preconditioner should result in smaller $k$ for the optimal trade-off).

In \Cref{fig:preconditioner_size} (left panel) we consider the aspirin dataset and vary the absolute kernel size $n$ .
Interestingly, we find that the convergence does not depend on the kernel size $n$. It is solely determined by the absolute number of inducing columns $k$. 
We observe a trade-off between iterative solver runtime and memory demand, i.e., for doubling the preconditioner the solver runtime halves, as can be seen by the linear relation in log-log space $\frac{\#}{n} \propto k^{-m}$ (see \Cref{fig:preconditioner_size}).

Next, in \Cref{fig:preconditioner_size} (right panel), we analyse this trade-off for different molecules and fixed kernel size $n=31\text{k}$.
We observe, that the same power-law relation persists for all molecules. 
However, offsets and slopes $m$ now depend on the molecule type, as expected based on their different spectra (see \Cref{app-sec:preconditioned_spectra}). 
We characterise the offset by the minimal preconditioner size $k_{\min}$, which indicates parity between iterative and closed-form solver runtime (intersection with solid line).
As a consequence, more complex molecules require a larger minimal preconditioner size $k_{\min}$ (see \Cref{tab:params_ruleofthumb}).

\subsubsection*{Deriving a Rule of Thumb}
We now discuss the subtleties of choosing the ideal number of inducing columns $k$, to minimize the overall runtime of the solver.
Here, we need to balance two aspects: Too few inducing columns lead to exceedingly long (effectively non-convergent) CG runtimes.
Secondly, too many inducing columns are prohibitively costly due to the quadratic scaling in $k$.
Hence, the optimal preconditioner size $k$ scales with the kernel size $n$, which means that using a fixed absolute (small or large) preconditioner size is clearly not a viable strategy. 
A simple heuristic is to use a fixed percentage of columns for preconditioning (e.g. $1\%$-baseline,  $k=\,\frac{n}{100}$)~\cite{chmiela2022iterativesgdml}.

We derive a more elaborate heuristic based on the overall computational costs of the preconditioned linear solver. These are given by the preconditioner costs $\mathcal{O}(k^2 n)$ plus the estimated iterative solver runtime (as formalised by the monomial relation), i.e., 
\begin{equation}    \label{eq:total_runtime_cost}
    \mathrm{runtime}[k] \propto n^3 \left[\left(\frac{k_{\min}}{k}\right)^m + \left(\frac{k}{n}\right)^2  \right]\, .
\end{equation}
Note, that we recover the original cubical runtime for ${k\to k_{\min}}$ and ${k\to n}$.
We then obtain a rule of thumb (RoT) to \emph{predict} the ideal number of columns $k^{\text{RoT}}$ for preconditioning via minimizing the overall cost:
\begin{equation}        \label{eq:ruleofthumb}
    k^{\text{RoT}} = \min_{k \in \left[1,\,n\right]}\Big\{\mathrm{runtime}\big[k\big]\Big\} = \left(\frac{\left(k_{\min}\right)^m m n^2}{2}\right)^{\frac{1}{2+m}}.
\end{equation}
As discussed in the beginning of this section, both hyperparameters ($k_{\min}$ and $m$) are connected to the spectral properties of the kernel matrix. Thus, they can be determined for smaller systems and later reused for larger systems. 
In \Cref{tab:params_ruleofthumb} we provide estimates for various molecules from the MD17 and MD22 datasets\cite{chmiela2017machine, chmiela2022iterativesgdml}. 
We observe, that the power-law trade-off coefficient $m$ is consistently close to unity, whereas the minimal preconditioner size $k_{\min}$ is associated with the size of the molecule and its complexity (dominant spectral properties).
We define default hyperparameters ($m=1$, $k_{\min} = 100$) and distinguish between a default vs. a (molecule)-specific RoT.
Here, the default RoT is directly applicable to new molecules/datasets, whereas the specific RoT show-casts the potential improvements via pre-training the hyperparameters.

\begin{table}[]
\centering
\def\arraystretch{1.3} 		
\setlength\tabcolsep{12pt}		
\caption{Rule of thumb hyperparameters ($m$ and $k_\text{min}$) measured with a kernel size of $n=31\text{k}$ (reusing experiments from \Cref{fig:preconditioner_size}). 
Here, $d$ is the number of atoms per molecule, $m$ represents the reduction in normalized iteration $\frac{\#}{n}$ associated with an increasing preconditioner size $k$. The minimal preconditioner size $k_{\text{min}}$ indicates equal runtimes between iterative and closed-form solve. The molecular conformation are taken from the MD17~\cite{chmiela2017machine} and MD22~\cite{chmiela2022iterativesgdml} datasets. 
}
\label{tab:params_ruleofthumb}
\begin{tabular}{lccc}
\toprule
\multicolumn{1}{c}{}        &  $d$ & $m$  & $k_{\min}$ \\ \cmidrule{1-4}
Default & ---                                 & 1    & 100        \\
Ethanol                     & 9                                   & 0.87 & 10         \\
Uracil                      & 12                                  & 1.07 & 32         \\
Toluene                     & 15                                  & 1.01 & 44         \\
Aspirin                     & 21                                  & 1.14 & 236        \\
Azobenzene                  & 24                                  & 1.02 & 62         \\
Catcher                     & 88                                  & 1.02 & 316        \\
Nanotube                    & 370                                 & 0.73 & 89         \\
\bottomrule
\end{tabular}
\end{table}

\subsubsection*{Experimental validation}
\begin{figure*}
    \centering
    \includegraphics[width=0.99\linewidth]{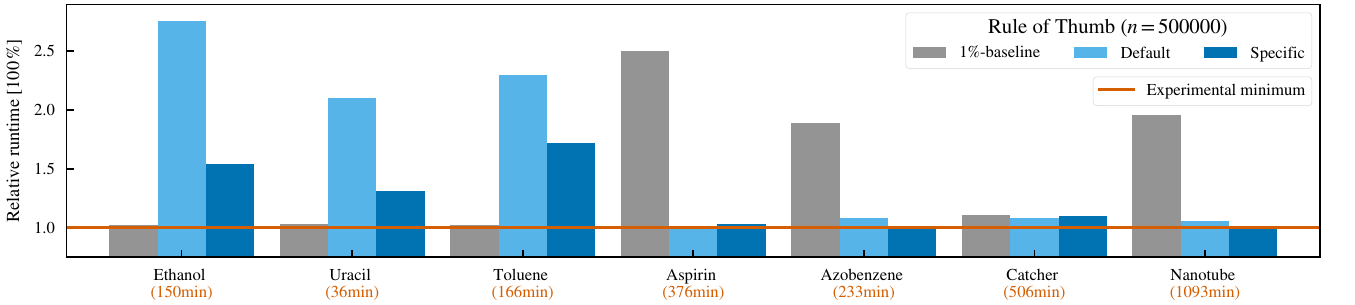}
    \caption{Minimizing the overall runtime by estimating the ideal number of inducing columns $k$ in advance. 
    All runtimes are given as relative percentage of experimental optimal runtime (orange), which is stated in parentheses. Fixed kernel size $n=500\text{k}$ (see \Cref{app-fig:ruleofthumb} for smaller kernel sizes).
    The rule of thumb is superior for complex and difficult to converge molecules, e.g. catcher or nanotube. For simpler and fast converging molecules (ethanol or uracil) a molecule specific rule of thumb can reduce the gap wrt. the $1\%$-baseline estimate ($k=\,\frac{n}{100}$). }
    \label{fig:ruleofthumb}
\end{figure*}

We validate our RoT by measuring the overall runtime depending on the number of inducing columns for various molecules.
Thereby, we obtain empirical estimates for the optimal number of inducing columns and the corresponding minimal runtime. 
We linearly interpolate all measurements to obtain runtime estimates for the default/specific RoT.

In \Cref{fig:ruleofthumb}, we show the relative runtime between the RoT prediction and the empirical minimal runtime. Additionally, we also compare to the simple $1\%$-baseline~(i.e. $k=\,\frac{n}{100}$)~\cite{chmiela2022iterativesgdml}. 
Interestingly, there are two distinct behaviors across our selection of datasets. Firstly, the small molecules (ethanol, uracil and toluene) for which the 1$\%$-baseline outperforms our RoT. 
Secondly, for larger and more complex molecules (aspirin, azobenzene, catcher and nanotube) our RoT is better than the $1\%$-baseline. 
Here, the $1\%$-baseline underestimates the preconditioner strength, which leads to significantly longer (potentially non-convergent) runtimes (plus 18h for the nanotube). 
Additionally, the RoT can be targeted to a specific molecule type. 
This further improves the performance of the specific RoT, as compared to its default counterpart see \Cref{fig:ruleofthumb}. 
In particular, for all simpler molecules the gap between $1\%$-baseline and specific RoT is reduced. 
Simultaneously, it retains the high performance for the more complex molecules. 

All findings are further verified by additional experiments for two smaller kernel sizes ($n=75\text{k}$ and $n=158\text{k}$) in \Cref{app-fig:ruleofthumb}.
Here, the specific RoT is even closer to the experimental minimum, since its molecule-specific parameters (\Cref{tab:params_ruleofthumb}) were estimated using a similar system size ($n=31\text{k}$).
This indicates that the specific RoT can be further improved through measuring the hyperparameters for a larger system, i.e., by reducing the gap between estimate and inference kernel size. However, obtaining these more accurate molecule-specific hyperparameters, increases the computational overhead.

\subsubsection*{Remarks on runtime and predictive accuracy}
So far, we have focused on algorithmic complexities, iterations times and relative solver runtimes in our theoretical considerations and experimental validation. 
We now investigate how transferable these consideration are to the overall solver, as this is what practitioners ultimately are interested in. 
To this end, we state the overall solver runtime (preconditioner construction + iterative solve) in the molecule label in orange in \Cref{fig:ruleofthumb}. For example, the training runtime of uracil is one order of magnitude lower in comparison to the nanotube.
This roughly agrees with their relative difference (based on CG iteration steps) in \Cref{fig:preconditioner_size} (right panel). 
In general, the computational ordering deduced from the number of iteration steps $\#$ in \Cref{fig:preconditioner_size} (right panel) align with the experimentally measured training runtimes (orange labels in \Cref{fig:ruleofthumb} and \Cref{tab:comparison_runtimeaccuracy}). This validates our previous analysis, which was mostly based on the iteration time. 
For example consider uracil and ethanol, which are consistently the simplest molecules to solve.
Alternatively, the most expensive molecules are the catcher and nanotube molecules. 
Note, the experimental nanotube runtime is twice as long as for the catcher molecule, which is, judging from their similar iterative runtimes in \Cref{fig:preconditioner_size}, somewhat unexpected. This effect is closely related to the costs of evaluating the kernel forward pass. Since the sGDML kernel marginalizes out all relevant symmetries~\cite{chmiela2019sgdml}, more symmetries increase the runtime complexity (28 [nanotube] > 4 [catcher]). This is also apparent when comparing the longer runtimes of ethanol (6 symmetries) with the faster training time of uracil (no symmetry).

Lastly, we compare the experimental runtimes and accuracies between a standard analytic (closed-form) solver and the preconditioned CG solver using the ideal number of inducing columns as predicted by our specific RoT.
The results in \Cref{tab:comparison_runtimeaccuracy} show no significant differences in the accuracy of the force prediction.
However, the iterative solver is up to $5x$ faster compared to its analytic (closed-form) counterpart and it uses significantly less memory.
We emphasize that the kernel size was limited to $n = 63\text{k}$ (matching 1k aspirin points) in our experiments, constrained by the analytic solver, which needs access to the complete kernel matrix ($\sim 30 \text{GB}$). In contrast, the preconditioned CG only requires to store a smaller preconditioner ($\sim 3 \text{GB}$) and thus allows to scale to much larger problem sizes with the same hardware (as done in \Cref{fig:ruleofthumb}). 
Note, that the cost for computing the preconditioner is negligible for a small number of inducing columns and hence preconditioned iterative solvers are effectively always superior to closed-form solvers for large systems \cite{Gardner2018_GPyTorch}. 

\begin{table}[]
    \centering
\caption{Comparing accuracy (mean-absolute error (MAE) of force prediction in $\text{mol }\text{kcal}^{-1} \text{ } \text{\r{A}}^{-1}$) and runtime between the analytic (closed-form) and preconditioned CG (PCG) solver for a fixed kernel size $n=63\text{k}$. Experiments were performed on a NVIDIA A100 (40GB). All runtimes are averaged across seven runs. The PCG performs indistinguishably at faster runtimes and lower memory demand.}
    \begin{tabular}{lccccc}
    \toprule
             &  \multicolumn{1}{c}{Accuracy}     &  \multicolumn{3}{c}{Runtime $[\text{min}]$} \\ 
    Molecule &  \tiny{$\Big|\text{MSE}_\text{PCG} - \text{MSE}_\text{Analytic}\Big|$} &  \small{Analytic} &  PCG  & Speed-up \\
    \midrule
Ethanol    &           0.00003 &             38 $\pm$ 4 &        10.5 $\pm$ 1.1     & $3.6\times$ \\
Uracil     &           0.00007 &              33 $\pm$ 6 &         6.5 $\pm$ 0.3   & $5.0\times$\\
Toluene    &           0.00002 &             41 $\pm$ 4 &       11.2 $\pm$ 0.8     & $3.6\times$\\
Aspirin    &           0.00006 &             35 $\pm$ 4 &        25.3 $\pm$ 1.6     & $1.4\times$\\
Azobenzene &           0.00012 &             35 $\pm$ 4 &       12.0 $\pm$ 0.9     & $2.9\times$\\
Catcher    &           0.00009 &             50 $\pm$ 5 &       50.4 $\pm$ 2.4     & $1.0\times$\\
Nanotube   &           0.00002 &            151 $\pm$ 17 &       70 $\pm$ 5        & $2.2\times$\\
    \bottomrule
    \label{tab:comparison_runtimeaccuracy}
    \end{tabular}
\end{table}

Overall, we conclude that our proposed heuristic provides an accurate and robust estimate for a practically well-performing number of inducing columns. This circumvents unnecessary large or too small preconditioners and can thus substantially speed-up training. 
Our default RoT provides reasonable estimates irrespective of molecule type and kernel size.
Moreover, we have demonstrated that our approach can be targeted to a specific molecule type and is therefore extendable to new molecules and datasets. 

\section{Conclusion}
To make large-scale kernel learning for MLFF reconstruction more widely accessible, this study reviews the combination of Nyström-type preconditioners and the iterative CG solver for model training. This approach is motivated by the following observation: Strong correlations between atoms give rise to exceedingly ill-conditioned loss surfaces, which can be readily addressed by selecting the appropriate inducing columns for a preconditioning step. The remaining stochasticity of the learning problem can then be effectively targeted with the Krylov subspace spanned by the iterative CG solver. In combination, both steps can be viewed as a low-rank expansion of the kernel matrix, which also allows for a chemically interpretation in terms of molecular fragments. 

We have demonstrated, how to construct preconditioners that enable super-linear convergence of the CG solver. This represents a significant speed-up over traditional closed-form approaches, in addition to the removal of their high memory demand, that we originally set out to alleviate. Remarkably, the effectiveness of the preconditioner scales with the number inducing columns $k$ and is largely independent of the kernel size $n$. This favorable scaling behavior is also a consequence of the aforementioned reoccurring correlation patterns between atoms, that stay largely the same across training points.

We have furthermore discussed important considerations when sampling inducing columns.
For simpler molecules, it suffices to sample uniformly, due to the strong correlation between individual training points (molecular geometries).
In contrast, more involved physical systems require a careful selection of inducing columns, as the relevant information is unevenly distributed. 
In that setting, basic sampling approaches are not sufficient as they lack the essential physical ingredients. This was the case in our catcher or nanotube examples, where the incomplete Cholesky decomposition was considerably more effective in finding a representative set of inducing columns, compared to the other approaches (except SVD).

Overall, we have shown how phyiscal insights about the problem at hand can be leveraged to build a more efficient training algorithm.
Moving forward, our findings can be combined with tools from automatic differentiation~\cite{schmitz2022algorithmic}, which allow to readily adapt models to new applications. This paves the way towards more challenging domains, such as biomedical molecules or materials, where it is even more instrumental to appropriately make use of powerful inductive biases.

Furthermore, our findings can be applied in the context of other recent advances in the field, such as the construction of more efficient descriptors~\cite{kabylda2022towards} or self-attention based learning~\cite{frank2022so3krates}. Both approaches allow global interactions, which can be potentially costly with a growing number of interacting atoms. Here, our understanding can help to mitigate some of the intrinsic scalability limitations towards larger system sizes.

\section*{Acknowledgements}
The authors have no conflicts to disclose.
SB, KRM, and SC acknowledge support by the German Federal Ministry of Education and Research (BMBF) for BIFOLD (01IS18037A). 
KRM was partly supported by the Institute of Information \& Communications Technology Planning \& Evaluation (IITP) grants funded by the government (MSIT) (No. 2019-0-00079, Artificial Intelligence Graduate School Program, Korea University and No. 2022-0-00984, Development of Artificial Intelligence Technology for Personalized Plug-and-Play Explanation and Verification of Explanation) and by the German Federal Ministry for Education and Research (BMBF) under Grants 01IS14013B-E, 01GQ1115.

\clearpage

\appendix

\renewcommand\thefigure{A\arabic{figure}}  
\setcounter{figure}{0}  

\renewcommand\thetable{A\arabic{table}}  
\setcounter{table}{0}

\section{Technical details} \label{app-sec:technical_details}
\SetKwInOut{KwIn}{Input}
\SetKwInOut{KwOut}{Output}
\subsection{SVD preconditioner}
The singular-value decomposition (SVD) provides access to the optimal low-rank approximation of $K$. Since $K$ is PSD the SVD is equivalent to the eigendecomposition, i.e. all left- and right-singular vectors are identical.
The SVD is given by
\begin{equation}
    K = U \, \Sigma \,\, U^{T} \, ,
\end{equation}
with ${U}$ being a unitary matrix consisting of all eigenvectors $\big[\pmb{u}_1, \ldots,\pmb{u}_n\big]$ and the diagonal matrix of singular values ${\Sigma} = \mathrm{diag}(\lambda_1, \ldots, \lambda_n)$.
An optimal rank-$k$ \emph{SVD preconditioner} is based on the $k$ dominant eigendirections, leading to the rank-$k$ approximation $K_k = U_k \, \Sigma_k \,\, U^{T}_k$. 
Importantly, this approach directly targets the dominant eigenvectors and explicitly removes their effect from the spectrum. In this sense, the \emph{SVD preconditioner} optimally flattens the kernel spectrum. 
Unfortunately, calculating the SVD scales quadratic with the kernel size $\mathcal{O}(k\,n^2)$ and therefore offers little practical value. However, for small system SVD preconditioning is computational feasible and we can use it as an theoretical lower bound for practical preconditioners. 

\textbf{Preconditioned CG as low-rank solver}
The solution of the linear system \Cref{eq:linear_system} can be expressed by the span of all non-trivial (i.e. up to the $\lambda$-regularized null space $l \approx 6 n_{\text{train}}$) kernel eigenvectors $\pmb{\alpha} \in \text{span}\left[\pmb{u}_0,\ldots, \pmb{u}_l\right]$. 
A CG solver expresses this solution iteratively within a Krylov basis $\mathcal{P} = \{p_1,\ldots,p_{\#}\}$, which approximates the complete spectrum simultaneously (hence $\text{span}[\mathcal{P}] \approx \text{span}[K_\lambda]$ after convergence). For example consider a degenerate flat spectrum (e.g. a fully preconditioned system), the CG converges in a single iterations, i.e. all $\left[\pmb{u}_1,\ldots, \pmb{u}_l\right]$ eigenvectors are summarized by a single search direction $\pmb{p}_1$. 
In contrast, for an ill-conditioned learning problem (non-uniform spectrum) the CG requires many iterations to converge. However, using $\# = l$ search directions we are guaranteed to recover the correct spectrum ($\text{span}[\mathcal{P}] = \text{span}[K_\lambda]$) and the approach becomes exact. 

We can now interpolate between both extremes using a preconditioner (\Cref{eq:preconditioned_linear_system}). 
A preconditioner can directly target the dominant eigenvectors and partially remove (pre-solve) the challenging physical eigenvectors. 
Hence, the resulting preconditioned linear system is 
characterised by $\text{span}[(K_k + \lambda I_n)^{-1} K_\lambda] = \text{span}[\pmb{u}_{k+1},\ldots, \pmb{u}_l] \approx \text{span}[\mathcal{P}]$ and overall the solution is expressed by two linear independent sets of basis vectors
\begin{equation}
    \pmb{\alpha} \in \text{span}[\pmb{u}_1, \ldots, \pmb{u}_k] + \text{span}[\pmb{p}_1,\ldots,\pmb{p}_{\#}].
\end{equation}
As we showed in \Cref{sec:experiment_rule_of_thumb}, it is possible to tune preconditioner size $k$ such that $k + \# \ll n$ and thereby construct a numerical efficient low-rank decomposition of the kernel.

\textbf{Atomic contribution} To intuitively visualize the low-rank decomposition defined by the SVD, we derive individual atomic contributions for each eigenvectors $\pmb{u}_i$. 
To this end, we reshape each $n$-dimensional eigenvector into its individual components: $n \mapsto (n_{\text{train}}, d, 3)$. 
Due to the periodicity within the structured eigenvectors (as discussed in \Cref{subsec:mlff_kernels}), we can average over the training dimension. Lastly, we measure the L2 distance wrt. the remaining three spatial dimensions to obtain a $d$-dimensional atomic contribution for each eigenvector. 

\subsection{Incomplete Cholesky decomposition} \label{app-sec:ICD}
To facilitate reproducibility, we provide pseudo code for an efficient incomplete Cholesky decomposition in Algorithm~\ref{app-alg:cholesky}. 
Importantly, this algorithm only requires access to pivot columns and the kernel trace. Additionally, it does not require to compute the full Schur complement $S$.
Thereby, it achieves minimal runtime $\mathcal{O}(k^2 \, n)$ and memory $\mathcal{O}(k \,n)$ cost.
\begin{algorithm}  
	\caption{incomplete Cholesky decomposition optimized for \textsc{numpy}, adapted from \cite{Harbrecht2012}}  \label{app-alg:cholesky}
	\KwIn{function: $\mathrm{get\_col}(i) \mapsto \pmb{k_i} \in \mathbb{R}^n$, \\ 
	array: $\pmb{d} := \mathrm{diag}(K)\in \mathbb{R}^n$, \\
	int: $k \leq n$ }
	\KwOut{array: $L_k:=[l_{i, m}] \in \mathbb{R}^{n \times k}$}
	\Begin{
		initialize $\pmb{\pi} :=(1, 2, \ldots, n)$ \\
        initialize $L_k := \texttt{zeros}(n, k)$        \\
		\For{$1 \leq m \leq k$ }{
			\tcp{Find pivot in permuted diagonal}
			set $i:= \mathrm{arg} \, \mathrm{max}\{d_{\pi_j}: j = m, m+1, \ldots, n\}$ \\
			swap $\pi_m$ and $\pi_i$ \\
			define $\pmb{\pi}_{>m}:=(\pi_j: j: m+1, \ldots, n)$\\
			\tcp{Set pivot and get new column}
			set $l_{\pi_m, m} := \sqrt{d_{\pi_m}}$ \\ 
			set $\pmb{k}_m = \mathrm{get\_col}(\pi_m)$ \\
			\tcp{Calculate Schur correction}
			\tcp{use np.einsum}
			$\pmb{s} := \sum_{j=1}^{m} l_{\pi_m, j} \cdot l_{\pmb{\pi}_{>m}, j}$ \tcp*[f]{len($\pmb{s}) = n - m$}\\
			
			set $l_{\pmb{\pi}_{>m}, m} := \left(k_{\pmb{\pi}_{>m}} - \pmb{s} \right) / l_{\pi_m, m}$ \\
			\tcp{Update diagonal}
			$d_{\pmb{\pi}_{>m}} = d_{\pmb{\pi}_{>m}} - \,\,l_{\pmb{\pi}_{>m}, m}^2$ 
		}
        \Return $L_k$ \\        
	}
\end{algorithm}

\section{Experimental results for more molecules} \label{app-sec:experiments_all_molecules}
The MD17 and MD22 datasets are available at \url{www.sgdml.org}.
The catcher dataset consists of a MD trajectory sampled at a temperature of 400 K and a resolution of 1 fs. Like in the other datasets, the potential energy and atomic force labels are calculated at PBE+MBD\cite{perdew1996generalized,tkatchenko2012accurate} level of theory. This dataset can be downloaded at \url{http://quantum-machine.org/datasets/catcher.npz}
\subsection{Preconditioned spectra} \label{app-sec:preconditioned_spectra}
\begin{figure}
    \centering
    	\def\arraystretch{0.9} 		
	\setlength\tabcolsep{1pt}		
    \begin{tabular}{c}
        \hspace{30pt}Aspirin, $n_\mathrm{train}=250$, $n=15750$           \\
        \includegraphics{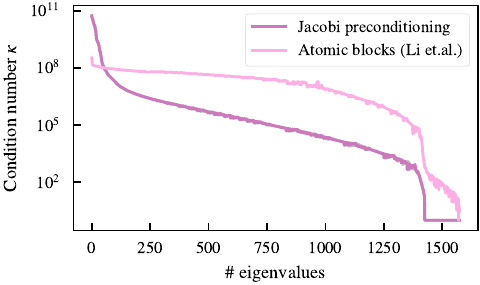}          
    \end{tabular}
    \caption{The effective spectrum of $P^{-1}K$ for simple Jacobi preconditioning and advanced atomic-block preconditioning \cite{li2021efficient}. The local structure prohibits to reduce the condition number sufficiently for practical usage as a preconditioner.}
    \label{app-fig:spectrum_local_preconditioning}
\end{figure}

\textbf{Jacobi preconditioning} is a simple and often effective alternative \cite{anzt2019adaptive,concus1985block,knyazev2001toward,lee2002performance, si2014memory}. It assumes a \emph{local} block-dominant matrix structure. In principle, it is therefore able to incorporate all structured prototype eigenvectors.
To experimentally show the necessity to incorporate global correlations, we show the effective spectra for local preconditioners in \Cref{app-fig:spectrum_local_preconditioning}. 
Firstly, we simply approximate the kernel matrix $K$ by a block-diagonal matrix (\emph{naive Jacobi} preconditioning). As expected, since the kernel matrix is not block dominant, this approach does not improve the conditioning. 
Secondly, we consider all correlations between identical atoms and zero out all other remaining structure. This generates a permuted block structure, which can be used as a platform for an advanced \emph{atomic-block} (Jacobi) preconditioning.
This physically motivated atomic-block approximation was proposed for approximative inference in \cite{li2021efficient}. 
However, this approach only mildly flattens the spectrum and is hence insufficient for practical preconditioning.  
This fundamentally relates to the fact that kernel matrices describe correlations between all training points, which can not be truncated.

Additionally, we provide preconditioned \textbf{spectra for more molecules} in \Cref{fig:eigvals}. As discussed in the main body of this manuscript, the incomplete Cholesky preforms nearly optimal for small number of inducing columns. However, for larger $k$ all Nyström-type preconditioner perform interchangeably.

\begin{figure*}
    \small
	\centering
	Ethanol, $n_\mathrm{train}=583$, $n=15741$
	\includegraphics[width=\linewidth]{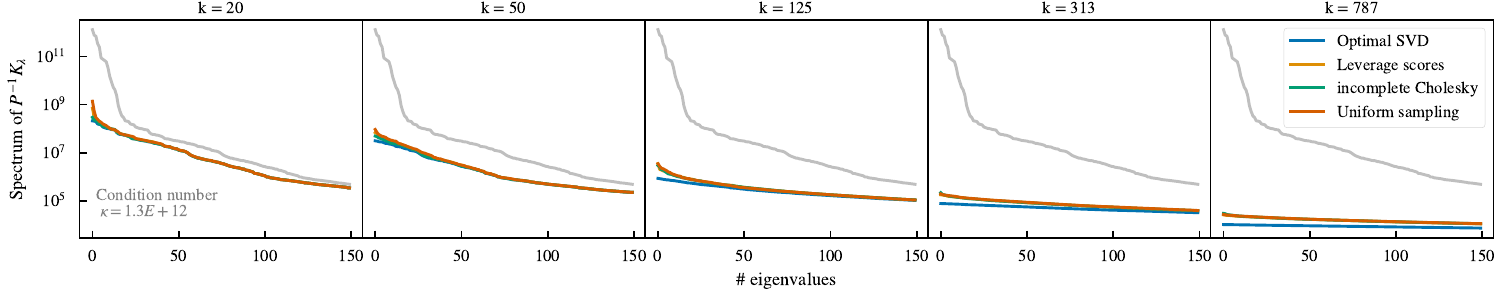}
	Uracil, $n_\mathrm{train}=437$, $n=15732$
	\includegraphics[width=\linewidth]{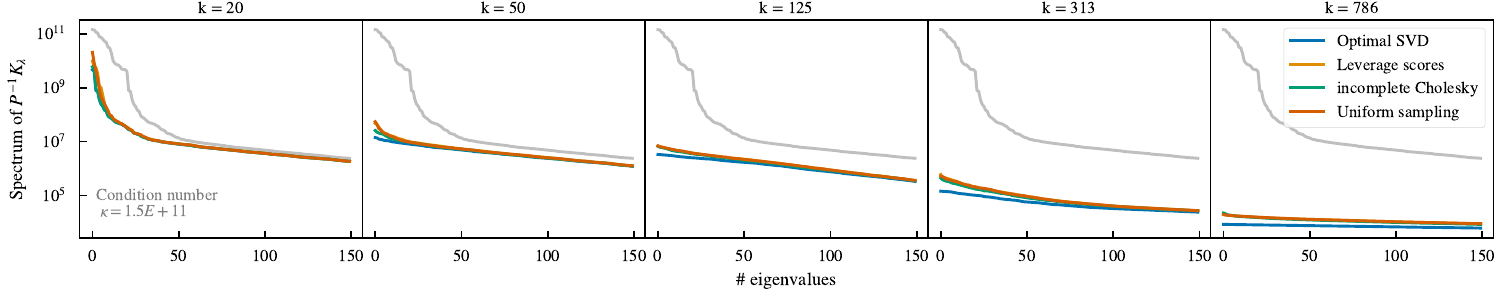}
	Toluene, $n_\mathrm{train}=350$, $n=15750$
	\includegraphics[width=\linewidth]{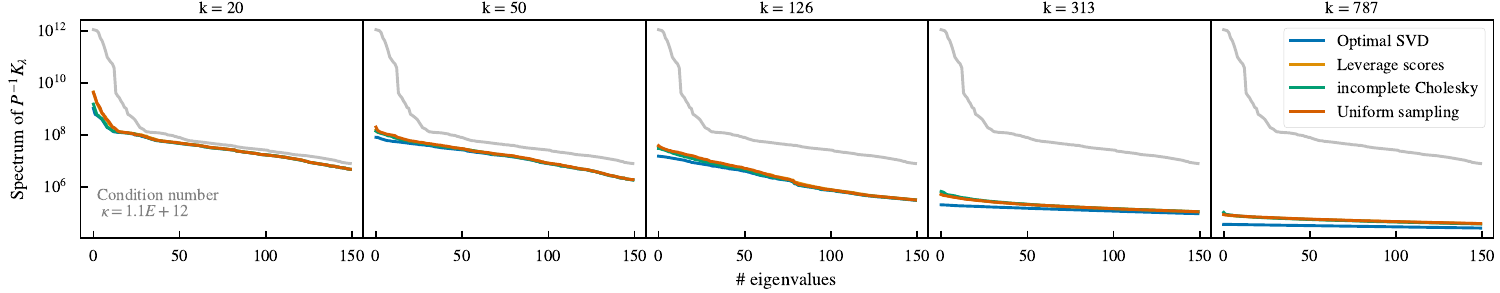}
	Azobenzene, $n_\mathrm{train}=202$, $n=14544$
	\includegraphics[width=\linewidth]{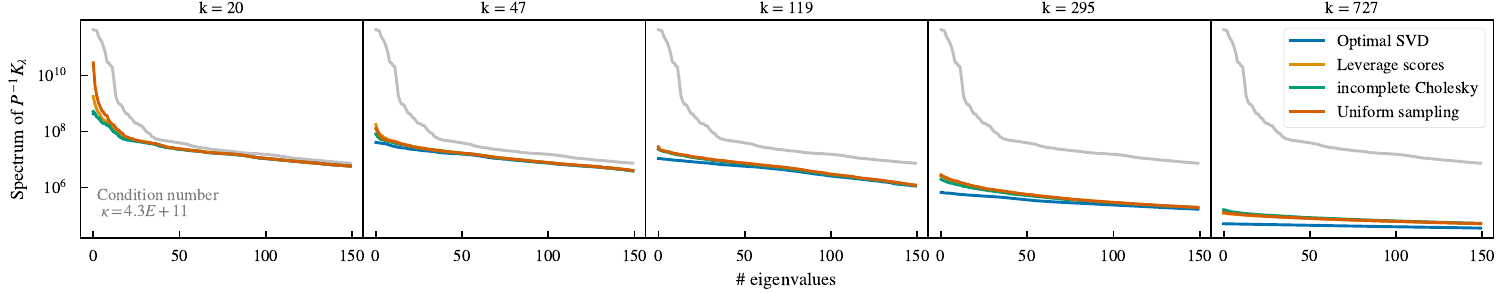}
	Catcher, $n_\mathrm{train}=59$, $n=15576$
	\includegraphics[width=\linewidth]{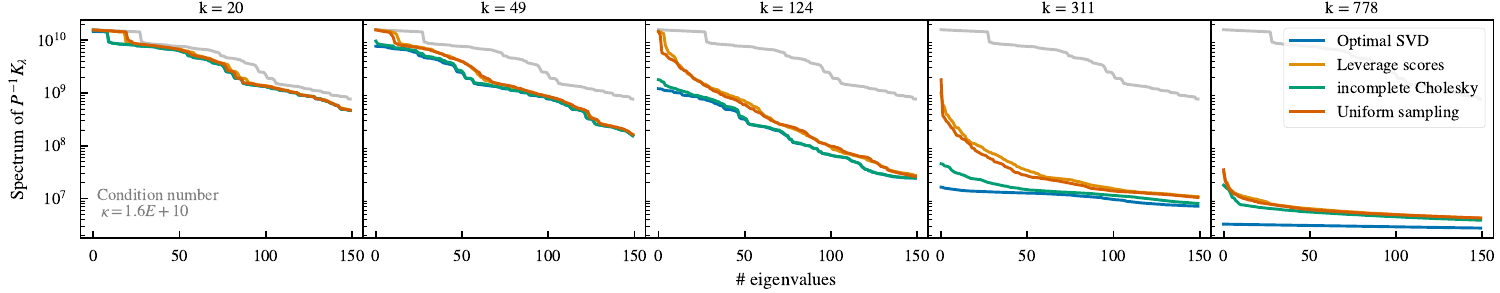}
    Nanotube, $n_\mathrm{train}=15$, $n=16650$
	\includegraphics[width=\linewidth]{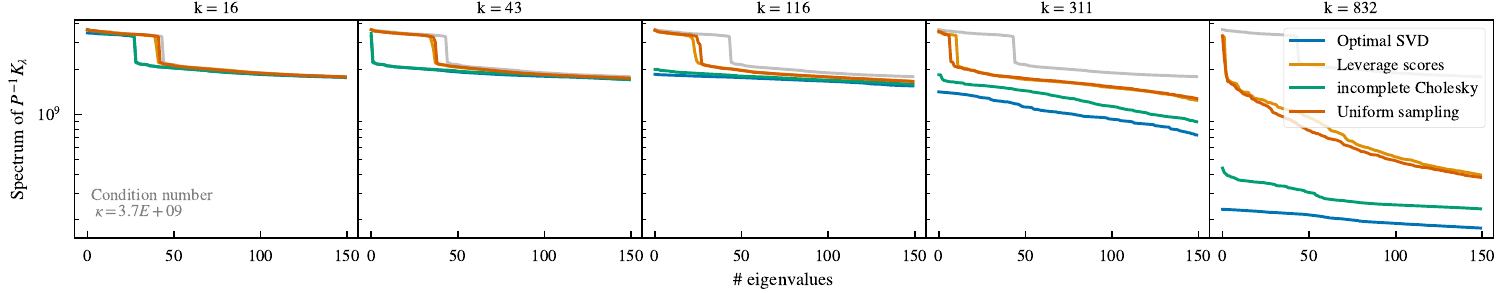}
	\caption{Preconditioned spectrum of $P^{-1}K$ for varying preconditioner size $k$ (restricted to the dominant 150 eigenvalues). }
	\label{app-fig:eigvals}
\end{figure*}

\subsection{Convergence of conjugate gradient} \label{app-sec:cg_convergence_all}
\begin{figure*}
    \centering
    \includegraphics[width=.99\linewidth]{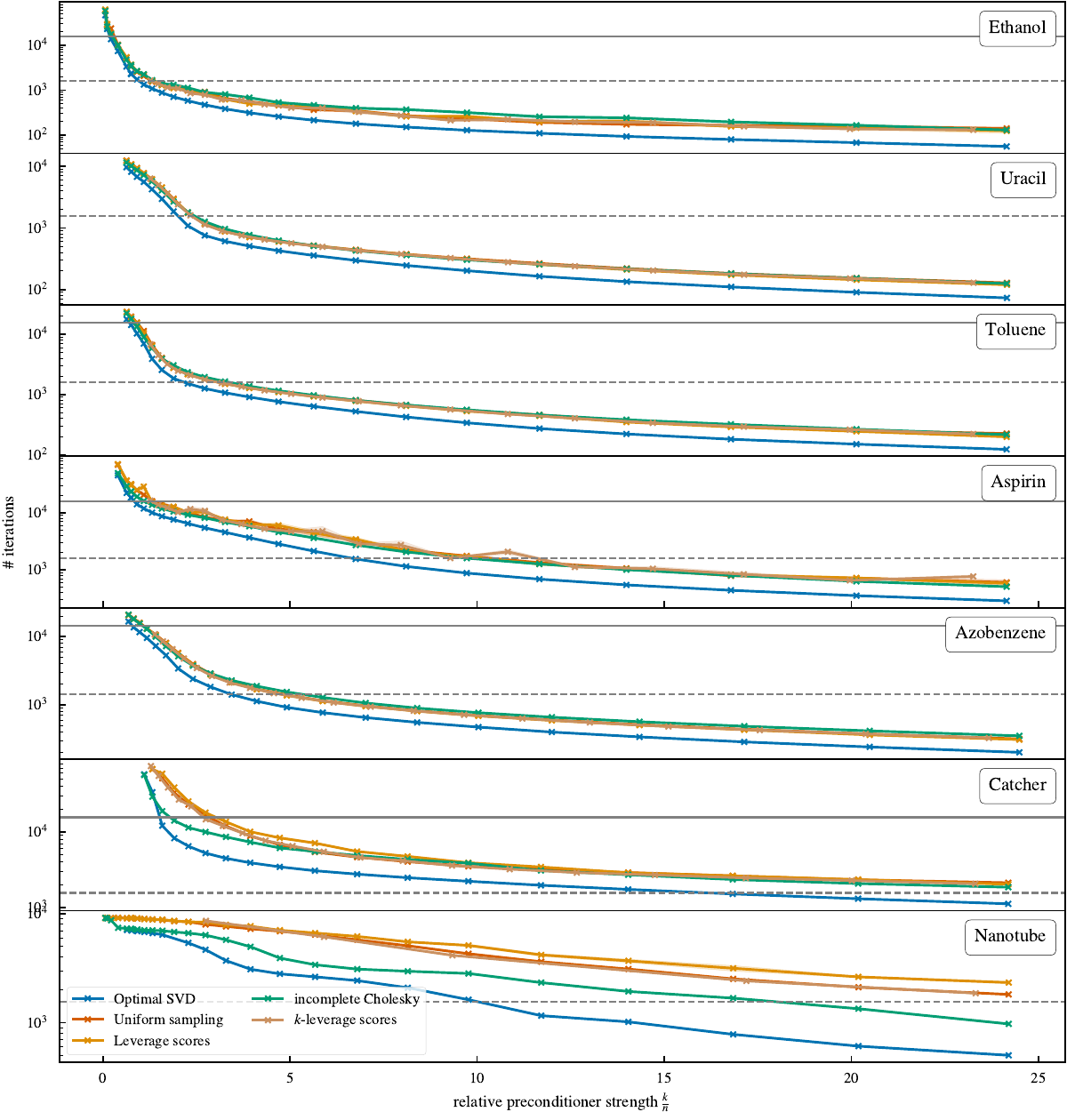}
 	\caption{Iterative convergence speed for varying preconditioner strength for fixed kernel size $n\approx 15\text{k}$ ($n_\text{train} = 250$ for aspirin). 
 Only minor performance differences are visible for ethanol. In contrast, for more complex molecules (catcher, nanotube) the incomplete Cholesky preconditioner is superior to naive uniform sampling of inducing columns.
 }
	\label{app-fig:cgsteps_all_molecules}
\end{figure*}

In \Cref{app-fig:cgsteps_all_molecules} we show the convergence for more molecules and further validate our findings described in the main text.  As previously observed, all Nyström-type preconditioners are similarly effective for most molecules and lower bounded the optimal SVD preconditioner. 

To further investigate the unexpected differences between uniform and leverage score based columns sampling, we consider an additional variant of the latter, namely rank-$k$ leverage scores~\cite{papailiopoulos2014provable}:
\begin{equation}
    \tau_i(K_k) = \big|[U_{k}]_{i,:}\big|_2^2.
\end{equation}
Here, $[U_k]_{i,:}$ denotes the $i$-th row of $U_k$. 
The sub-dominant spectral directions are diminish via omitting the corresponding eigenvectors. This allows to directly match the rank-$k$ for the leverage scores estimation to the preconditioner size $k$. Thereby, the rank-$k$ leverage scores adapt to the changing dimensionality of preconditioner subspace.
This is in contrast to ridge leverage scores: Here, sub-dominant spectral dimensions were suppressed by a regularization $\lambda$, which was matched to the overall ridge regularization from \Cref{eq:linear_system}. Thereby, ridge leverage scores are indicative for a fixed regularized kernel spectrum.
Lastly, we emphasize that this definition of rank-$k$ leverage scores requires to calculate the SVD decomposition and is therefore too expensive for practical usability. One possibility would be to use ridge leverage scores with an adaptive regularization $\lambda$ as induced by the requested preconditioner size $k$~\cite{mccurdy2018ridge}.

The experimental results in \Cref{app-fig:cgsteps_all_molecules} show that the improved rank-$k$ leverage scores bridge the gap to the performance of uniform sampling, as observed for the catcher and nanotube molecule. However, there are no significant differences between both approaches and both are clearly lower-bounded by the incomplete Cholesky column selection strategy.

\clearpage
\section{Rule of thumb} \label{app-sec:rule_of_thumb}
\begin{figure}[h!]
    \centering
    \includegraphics[width=0.96\linewidth]{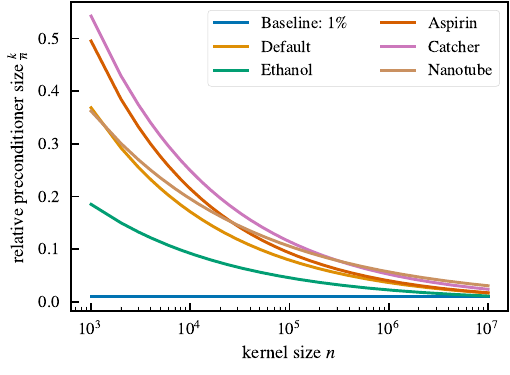}
    \caption{Comparing the predicted preconditioner size of all used heuristics depending on the kernel size $n$. }
    \label{app-fig:prediction_ruleofthumb}
\end{figure}
We experimentally validate our heuristic rule of thumb for two additional, smaller kernel sizes in \Cref{fig:ruleofthumb}. To corresponding hardware specification are summarized in \Cref{app-tab:RoT_hardware_specs}. 
For the readers convenience we compare the prediction for all heuristics (default/specific RoT, naive baseline) depending on the kernel size $n$ in \Cref{app-fig:prediction_ruleofthumb}. In contrast to the naive baseline, our RoT adapts to the kernel size $n$ and is therefore applicable over several orders of magnitude.

\begin{figure*}
    \centering
    \includegraphics[width=0.99\linewidth]{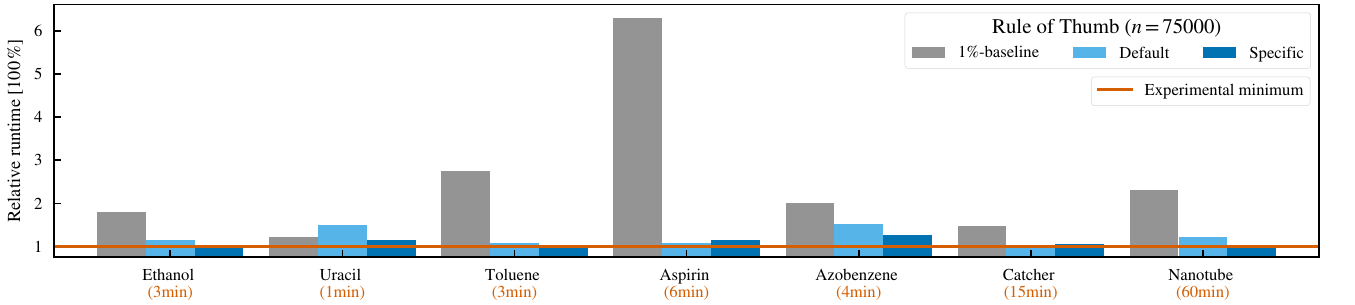}
    \includegraphics[width=0.99\linewidth]{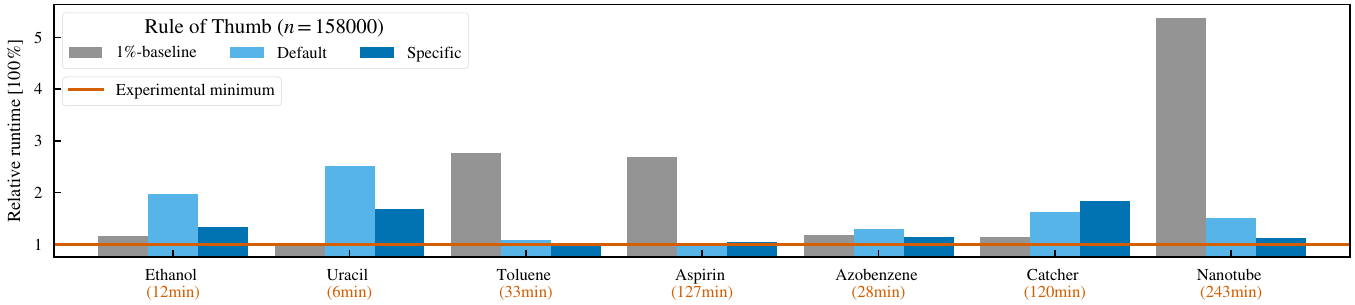}
    \caption{Experimental validation of the rule of thumb for various molecules and fixed kernel size $n=75\text{k}$ and $n=158\text{k}$. The minimal runtime is printed in parentheses.}
    \label{app-fig:ruleofthumb}
\end{figure*}

\begin{table*}
\caption{Hardware specification used for the RoT validation experiments (\Cref{fig:ruleofthumb}~and~\ref{fig:ruleofthumb}).}
\label{app-tab:RoT_hardware_specs}
\begin{tabular}{llllll}
\toprule
 $n$      &     &                 Ethanol &                  Uracil &      Toluene &                 Aspirin \\
\midrule
75000  & CPU &              Gold 6226R &              Gold 6226R &   Gold 6226R &              Gold 6226R \\
       & GPU &  Quadro RTX 6000 (24GB) &  Quadro RTX 6000 (24GB) &  A100 (40GB) &  Quadro RTX 6000 (24GB) \\
157500 & CPU &              Gold 6226R &              Gold 6226R &   E5-2690 v4 &              E5-2690 v4 \\
       & GPU &             A100 (40GB) &             A100 (40GB) &  P100 (12GB) &             P100 (12GB) \\
504000 & CPU &              Gold 6226R &              Gold 6226R &   Gold 6226R &              Gold 6226R \\
       & GPU &             A100 (40GB) &  Quadro RTX 6000 (24GB) &  A100 (40GB) &  Quadro RTX 6000 (24GB) \\
\midrule
\end{tabular}
\begin{tabular}{lllll}
   $n$    &     &              Azobenzene &      Catcher &                Nanotube \\
\midrule
75000  & CPU &              Gold 6226R &   Gold 6226R &              E5-2650 v4 \\
       & GPU &  Quadro RTX 6000 (24GB) &  A100 (80GB) &        TITAN RTX (24GB) \\
157500 & CPU &              Gold 6226R &   E5-2690 v4 &              Gold 6226R \\
       & GPU &             A100 (40GB) &  P100 (12GB) &             A100 (40GB) \\
504000 & CPU &              Gold 6226R &   Gold 6226R &              Gold 6226R \\
       & GPU &  Quadro RTX 6000 (24GB) &  A100 (40GB) &  Quadro RTX 6000 (24GB) \\
\bottomrule
\end{tabular}
\end{table*}

\clearpage

\bibliographystyle{achemso}
\bibliography{bibfile}

\end{document}